\documentclass[pra,twocolumn,showkeys]{revtex4}
\usepackage{bbm}
\usepackage{natbib}
\usepackage{graphics}
\usepackage{amsmath}
\usepackage{mathrsfs}
\usepackage{theorem}
\usepackage{float}
\usepackage{hyperref}
\usepackage{rotating}
\usepackage{amssymb}
\usepackage[english]{babel}
\usepackage{color}
\usepackage{CJK}
\usepackage{multirow}
\usepackage{lscape}
\usepackage{fancybox}

\newcommand{\ket}[1]{|#1\rangle}
\newcommand{\bra}[1]{\langle #1|}

\newcommand{\be}{\begin{eqnarray}}
\newcommand{\ee}{\end{eqnarray}}

\begin{document}
\title{Open dynamics in the Aubry-Andr\'{e}-Harper model coupled to a finite bath: the influence of localization in the system and  dimensionality of bath}
\author{H. T. Cui $^{1, 2}$}
\email{cuiht01335@aliyun.com}
\author{M. Qin $^{1}$}
\author{L. Tang $^{1}$}
\author{H. Z. Shen $^{2}$}
\email{shenhz458@nenu.edu.cn}
\author{X. X. Yi $^{2}$}
\email{yixx@nenu.edu.cn}
\affiliation{$^1$ School of Physics and Optoelectronic Engineering, Ludong University, Yantai 264025, China}
\affiliation{$^2$ Center for  Quantum Sciences, Northeast Normal University, Changchun 130024, China}
\date{\today}

\begin{abstract}
The population evolution of single excitation is studied in the  Aubry- Andr\'{e}- Harper (AAH) model coupled to a $d (=1,2,3)$-dimensional simple lattices bath with a focus on the effect of localization in the system and the dimensionality of bath. By performing a precise evaluation of time-independent Schr\"{o}dinger equation, the reduced energy levels of the system can be determined. It is found that  the reduce energy levels show significant relevance for the bath dimensions. Subsequently, the time evolution of excitation is studied in both the system and bath.  It is found that excitation  in the system can  decay  super-exponentially when $d=1$ or  exponentially  when $d=2,3$. Regarding the finite nature of bath, the spreading of excitation in the lattices bath is also studied. We find that, depending on the dimensions of bath and the initial state, the spreading of excitation in the bath is diffusive or behaves localization.
\end{abstract}


\maketitle

\section{introduction}

A surge of interest has been observed  in recent years regarding the dynamics of open many-body systems. The reasons for this can be summarized as the follows. First, since no quantum system in a  laboratory setting  can be excluded from the influence of the environment, it  is important to establish how  quantum coherence in a system decays. Furthermore, because of the ubiquitous  correlations and complex interactions in many-body systems, decoherence is expected to display different features compared with the few-body quantum systems.   It is thus urgent  that  the special role of  correlations and interaction upon the open dynamics of many-body systems be identified. Second, by manipulating the coupling to the environment, the quantum coherence in a system can be precisely engineered \cite{diehl08, verstraete09}. Consequently, through the quantum-environment engineering,  the quantum computation can be implemented \cite{verstraete09}, and the many-body effect can be simulated in a controllable manner \cite{gan14}. Finally, open many-body systems can exhibit a wide range of nonequilibrium  features that not found in equilibrium systems, e.g.  discrete time crystals \cite{yao17} and dissipative phase transition \cite{dps}. In these situations, the environment plays a crucial role.

Recently, the open dynamics of   localization-delocalization transition systems have received significant attention because of the experimental realization of  many-body localization transition (MBL) in ultra-cold atoms \cite{exp-quasidisorder}. In experiments,  the  atom-atom collisions and imperfect trapping were indicated as important influences  on the observation of MBL \cite{luschen2017}. Additionally,  the theoretical studies  reveal  that although  localization may be destroyed  because of   coupling to a thermalizing  environment, there is a large time window during which the localization can be preserved  \cite{opendisorder, luschen2017}. Moreover,  localization in the system can impose its influence on the finite environment \cite{quantumbath}.  Consequently,   the localization of the environment can be changed. This is known as the proximity effect \cite{proximity}, and the dynamics in such a  system will inevitably be non-Markovian.

As for the existing statement, it is useful to establish the interplay between the localization in a system and decoherence induced by coupling to the structured environment. The word ''structured"  in this instance refer to the environment that  displays a nontrivial band structure, or a setting with a  finite  spectrum. This consideration is  reasonable from an experimental perspective. For example, in the cold atomic gases  in optical lattices, three pairs of counterpropagating laser beams  are imposed in  orthogonal directions  to perfectly control the motion of atoms and the interactions among them \cite{ol}. Accordingly,  the imposed laser field  corresponds to a highly dimensional photon bath. A similar situation can also occur for cold atoms in  photonic crystal (PC) lattices. Based on the dimensionality and the exotic band structure in  PC lattices,   atom-atom interactions can be mediated using a versatile set of approaches \cite{douglas2015, tudela2015, tudela2018}. An important  consequence linked to the presence of a structured environment is the failure of the Born-Markov approximation. As such, non-Markovianity is expected to have an important impact on the open dynamics of the localized systems.

In this work,  we aim to derive the decaying dynamics of single excitation in  a structured environment by focusing on the influence of  localization of the system, the dimensionality of the environment, and the initial conditions. To achieve this goal, the population evolution of single excitation in both  a one-dimensional atomic chain (the system) and  a $d (=1, 2, 3)$-dimensional simple lattices bath is studied precisely. We find that although   localization in the atomic chain can eventually be destroyed, the  excitation population can indicate distinct properties in both  the system and the lattices bath, respective of the  localization in the initial state and  the dimensionality of bath.  The following discussion is divided into four sections. In section II, the Hamiltonians of our model are presented. Following on,  the reduced energy levels in the system are analytically evaluated. In section III, the population evolution of single excitation is studied in both atom chains and lattices bath, focusing on the localization of the system, the initial state and the dimensionality of the bath. Finally section IV presents additional discussion and a conclusion.

\section{The  model and reduced energy levels of the system}

We focused on  a one-dimensional tight-binding  atomic chain with onsite modulation (the system), coupled to a  $d$-dimensional simple lattices bath. The Hamiltonian can be written as
\be
H=H_s +H_b + H_I,
\ee
with
\be
H_s&=&\sum_{n=1}^{N_s} \lambda\left(a^{\dagger}_{n-1} a_{n} +a^{\dagger}_n a_{n-1} \right) + \Delta\cos(2\pi \beta n +\phi)a^{\dagger}_n a_{n},\nonumber\\
H_b&=& - J\sum_{\langle\mathbf{i, j}\rangle}  b^{\dagger}_{\mathbf{i}} b_{\mathbf{j}} + H.c., \nonumber\\
H_I&=& g \sum_{n} a^{\dagger}_n b_{\mathbf{r}_n} + H.c., \nonumber
\ee
where $N_s$ denotes the length of the atomic chain, and $a_n (a^{\dagger}_n)$ is the annihilation (creation) operator of excitation at the $n$-th atom.  $H_s$ characterizes the  Aubry-Andr\'{e}-Harper (AAH)  model when $\beta$ is a Diophantine number \cite{jitom1999}. It is known for AAH model that there is two distinct phases, i.e., the  delocalized  ($\Delta/\lambda<2$) and the localized phase ($\Delta/\lambda>2$), in which the eigenstate of $H_S$ behaves extended or localized in the position space \cite{aah}. Moreover, AAH model  resembles to the two-dimensional quantum Hall system \cite{kraus}. Thus a topological in-gap edge mode can also found, in which the excitation becomes localized at the boundary \cite{kraus}. Moreover, the AAH model can be realized in cold atomic gases, and  experimental exploration of the  delocalization-localization phase transition in the AAH model  has been  extensively pursued \cite{exp-quasidisorder}. Throughout this paper, $\beta=\left(1+\sqrt{5}\right)/2$ is selected. In this context, the system is quasi-periodic or quasi-disordered. Furthermore, the localized edge mode can occur in the system under open boundary conditions, which is topologically nontrivial  and occurs as an in-gap energy level \cite{lang2012}. As shown in the following discussion, the edge mode will have an important effect on the excitation  population in the atomic chain. Finally, it is commented that a different choice of $\beta$ is possible only if $\beta$ is a irrational number. The physics of AAH model does not depend on the special value of $\beta$.

$H_b$ depicts a $d$-dimensional simple lattice bath with nearest-neighbor hopping. Under periodic boundary conditions, $H_b$ can be diagonalized in momentum space as follows,
\be\label{hb}
H_b=- \sum_{\mathbf{k}} \omega(\mathbf{k}) b^{\dagger}_{\mathbf{k} } b_{\mathbf{k} },
\ee
where $\omega(\mathbf{k})=2J\sum_{q=1}^d \cos k_q$ and $b_k (b_k^{\dagger})$ is the bosonic annihilation (creation) operator of the $k$-th  mode. It is convenient to restrict  $k_q\in\left[-\pi, \pi\right] \left(q=1,2, \cdots, d\right)$, and consequently $b_k=1/\left(2\pi\right)^{d/2}\sum_{\mathbf{j}}e^{- i \mathbf{k}\cdot \mathbf{j}}b_{\mathbf{j}}$, and $\omega(k)/2J \in \left[-d, d\right]$. $H_I$ depicts the local coupling, that the atomic site $n$ is coupled uniquely  to the nearest neighbored  lattice site $\mathbf{r}_n$. In momentum space, it can be transformed  into
\be
H_I= \frac{g}{\left(2\pi\right)^{d/2}}\sum_{n, \mathbf{k}}  e^{i \mathbf{k} \cdot \mathbf{r}_n}b_k a^{\dagger}_n +H. c.,
\ee
where $\mathbf{r}_n$ denotes the position vector of the $n$-th atomic site in the  real space of $d$-dimensional lattice bath. Coupling strength $g$ is assumed to be homogeneous for any wave vector $\mathbf{k}=\left(k_1, k_2, \cdots, k_d\right)$. Finally, it should be noted that the model can be readily realized  experimentally, e.g., in PC lattices \cite{douglas2015, tudela2015}.

Although  atomic interaction is of significant importance for observing MBL transition  \cite{exp-quasidisorder}, the current research does not appear intent on including this aspect since atomic interaction will render the population evolution of excitation too complex in both the system and bath.   As such,  it is convenient to restrict our discussion to a single excitation case. Thus, the eigenstate for the total hamiltonian $H$ can be written formally  as
\be\label{psie}
\ket{\psi_E}&=& \left(\sum_{n=1}^{N_s} \alpha_n \ket{1}_n \right)\otimes \ket{0}_{\mathbf{k}} + \nonumber \\ &&\ket{0}^{\otimes N_S} \otimes \left(\sum_{\mathbf{k}} \beta_{\mathbf{k}} \ket{1}_{\mathbf{k}}\right),
\ee
in which $\ket{1}_n = a_n^{\dagger}\ket{0}$ denotes the occupation of the $n$-th atomic site, $\ket{0}_k$ is the vacuum state of $b_k$, and $\ket{1}_{\mathbf{k}}=b_{\mathbf{k}}^{\dagger}\ket{0}_{\mathbf{k}}$. Substituting $\ket{\psi_E}$ into Schr\"{o}dinger equation $H\ket{\psi_E}= E \ket{\psi_E}$, the following is obtained $(\hbar \equiv 1)$
\addtocounter{equation}{1}
\begin{align}\label{seqa}
\lambda\left(\alpha_{n+1} + \alpha_{n-1}\right) +\Delta \cos(2\pi \beta n +\phi)\alpha_n +\nonumber\\ g  \sum_{\mathbf{k}} e^{i \mathbf{k}\cdot \mathbf{r}_n} \beta_{\mathbf{k}} =&E \alpha_n;  \tag{\theequation a} \\
-\omega(\mathbf{k}) \beta_{\mathbf{k}} +  \frac{g}{\left(2\pi\right)^{d/2}}\sum_{j=1}^{N_s} \alpha_j e^{- i \mathbf{k} \cdot \mathbf{r}_j}=&E \beta_{\mathbf{k}}. &&\tag{\theequation b}
\label{seqb}
\end{align}
After solving the Eq.  (\ref{seqb}), one gets
\be
\beta_{\mathbf{k}}=\frac{g}{\left(2\pi\right)^{d/2}}\frac{\sum_j \alpha_j e^{- i \mathbf{k} \cdot \mathbf{r}_j} }{E+\omega(\mathbf{k})}.
\ee
Substituting this expression into Eq. (\ref{seqa}) and replacing $\sum_{\mathbf{k}}$ by its integral, one obtains
\be\label{seq}
&&\lambda\left(\alpha_{n+1} + \alpha_{n-1}\right) +\Delta \cos(2\pi \beta n +\phi)\alpha_n +\frac{g^2}{\left(2\pi\right)^d}\times \nonumber \\
&&\sum_{j=1}^{N_s} \alpha_j \int_{-\pi}^{\pi}\text{d}k_1 \int_{-\pi}^{\pi}\text{d}k_2\cdots \int_{-\pi}^{\pi}\text{d}k_d\frac{e^{ i \mathbf{k} \cdot \left(\mathbf{r}_n- \mathbf{r}_j\right)}}{E + \omega(\mathbf{k})}= E \alpha_n,
\ee

Eq. \eqref{seq}  constitutes a linear system of equations for $\alpha_n$. The eigenenergy, i.e., $E$, corresponds to the zero  determinant of the coefficient matrix. It is noted that the degree of freedom of bath has been eliminated in the above derivation. For clarity, in the following discussion, Eq. \eqref{seq}  is referred  as the reduced  Schr\"{o}dinger equation of the system, and $E$ is referred as the reduced eigenenergy of system. The integral, i.e.,
\be\label{int}
f_d\left(n\right)= \frac{1}{\left(2\pi\right)^d} \int_{-\pi}^{\pi}\text{d}k_1 \int_{-\pi}^{\pi}\text{d}k_2\cdots \int_{-\pi}^{\pi}\text{d}k_d\frac{e^{ i \mathbf{k} \cdot \mathbf{r}_n}}{E + \omega(\mathbf{k})}.
\ee
is known as the lattice Green function. $f_d\left(n\right)$ is symmetric under the reflection $\mathbf{r}_n \leftrightarrow -\mathbf{r}_n$.  Because of the existence of denominator $E + \omega(\mathbf{k})$,  the evaluation of $f_d\left(n\right)$ has to be implemented for two distinct situations. One is for $\left|E\right| > \max_{\mathbf{k}} \left|\omega(k)\right|$, whre there is no singularity in $f_d\left(n\right)$. Thus, it can be determined in numerics.  In the other point, when $\left|E\right| < \max_{\mathbf{k}} \left|\omega(\mathbf{k})\right|$, $f_d\left(n\right)$ is  divergent because of the occurrence of poles decided by  $E + \omega(\mathbf{k})=0$. To encounter this difficulty, one  has to extend $f_d\left(n\right)$ into the complex $E$ space. Thus, $f_d(n)$ is transformed into a contour integral with a detour around the poles, which can be evaluated using residue methods. As  a consequence, the hermiticity of coefficient matrix is broken, and thus the complex $E$ can occur, which characterizes the decaying of excitation into the bath. In physics, the divergence of the integral is meaningless. In order to find physically reasonable  results, one has to introduce complex $E$ to eliminate this divergence. This method has been extensively used in the quantum scattering theory to eliminate the divergence induced by the fact that the scattered particle could move infinitely\cite{weinberg}.

It is convenient to assume for the atomic sites to be arranged along the lattice direction of the bath when  $d=1$. While, when $d=2, 3$,  the atomic chain is arranged along the body-diagonal direction of the bath  such that $e^{ik \mathbf{r}_n}$ is isotropic in all directions. This selection is not only beneficial  to the evaluation of $f_d(n)$, but also convenient to display the influence of dimensionality of bath. Admittedly, a different arrangement of atomic chain will inevitably change the form of $e^{ i \mathbf{k} \cdot \mathbf{r}_n}$. However, since $f_d(n)$ is determined mainly by its poles decided by the relation $E + \omega(\mathbf{k})=0$, this change will have no intrinsic affect on the evaluation of $f_d(n)$. Thus, it is anticipated that the conclusion does not depend on the special choice of the arrangement.  By the results in \cite{ray14} and \cite{jd04}, one obtains
\be\label{f1}
f_1\left(n\right)&=&\frac{\pi }{\sqrt{e^2-1}} \frac{\text{sign}(e)}{\left(\left|e\right|+\sqrt{e^2 -1}\right)^{n}};\\
\label{f2}
f_2\left(n\right)&=& -\frac{16 \cos n\pi}{\pi^2 (4 n^2 -1)e^2} \times\nonumber \\
&&F\left(1, 1, 1, \tfrac{3}{2}, \tfrac{3}{2};  \tfrac{3}{2} + n, \tfrac{3}{2} -n, \tfrac{3}{2}, \tfrac{3}{2}; \tfrac{4}{e^2}\right);\\
\label{f3}
f_3\left(n\right)&=& \frac{(-1)^n}{e} \frac{\left(3n\right)!}{\left(3^n n!\right)^3}\left[\frac{e}{3}\left(1- \sqrt{1- \frac{9}{e^2}}\right)\right]^{3n}\times\nonumber\\
&&F\left(\tfrac{1}{3}, \tfrac{2}{3}; n+1; \eta_+\right)F\left(\tfrac{1}{3}, \tfrac{2}{3}; n+1; \eta_-\right),
\ee
in which $e=E/2J$.  $\eta_{\pm}= \tfrac{1}{8e^2}\left[4e^2 + \left(9-4e^2\right)\sqrt{1- \tfrac{9}{e^2}}  \pm 27 \sqrt{1- \tfrac{1}{e^2}}\right]$. $F\left( x_1, x_2, \cdots; y_1, y_2, \cdots; z \right)$ is the  hypergeometric function, which converges for $\left|z\right|<1$. Notably,  $f_d(n)$ is absolutely divergent  at the boundary $\left|e\right|=\pm d$. These special cases were excluded in the remaining analysis of this section.

\begin{figure*}[t]
\center
\includegraphics[width=12cm]{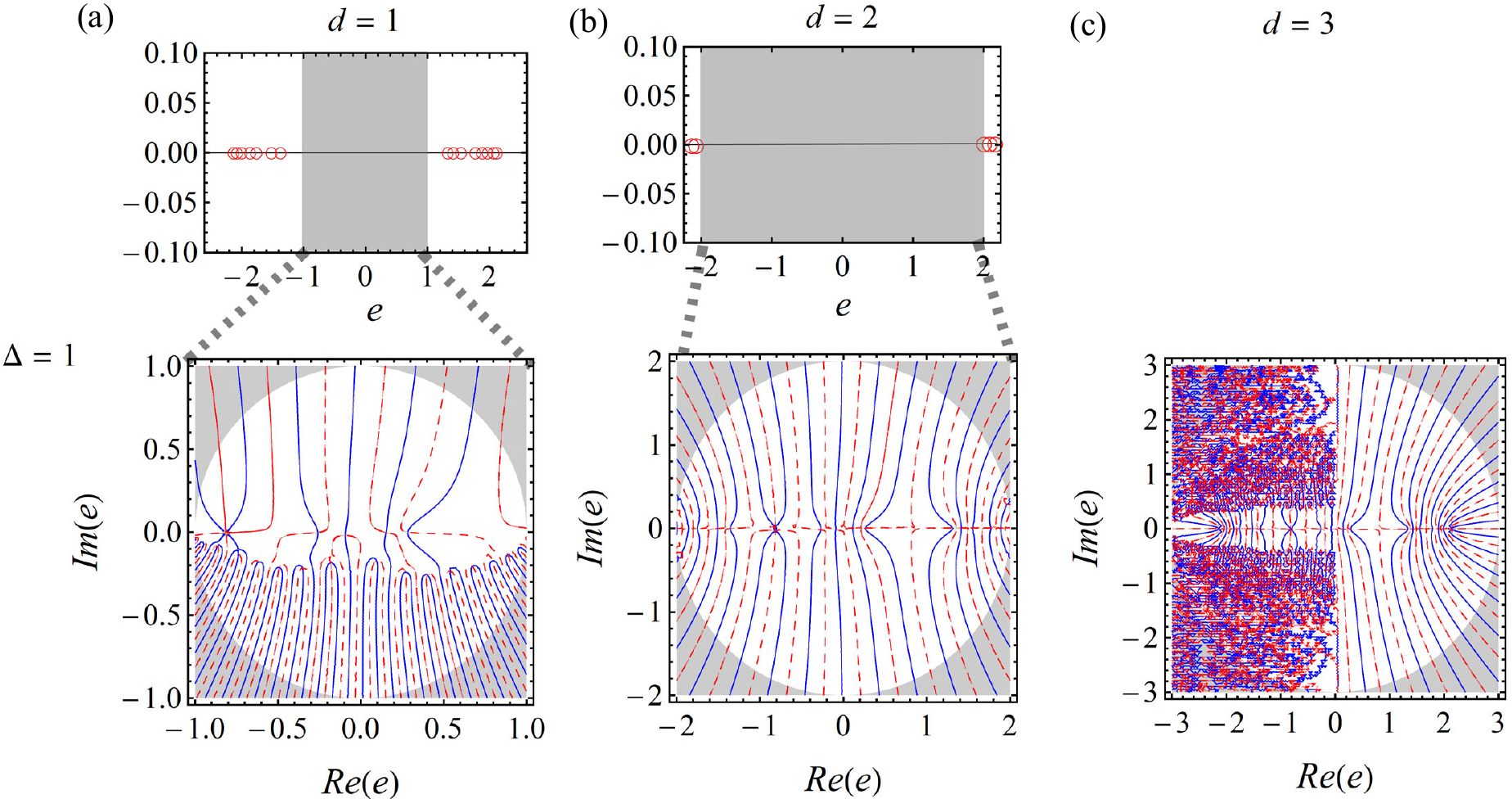}
\includegraphics[width=12cm]{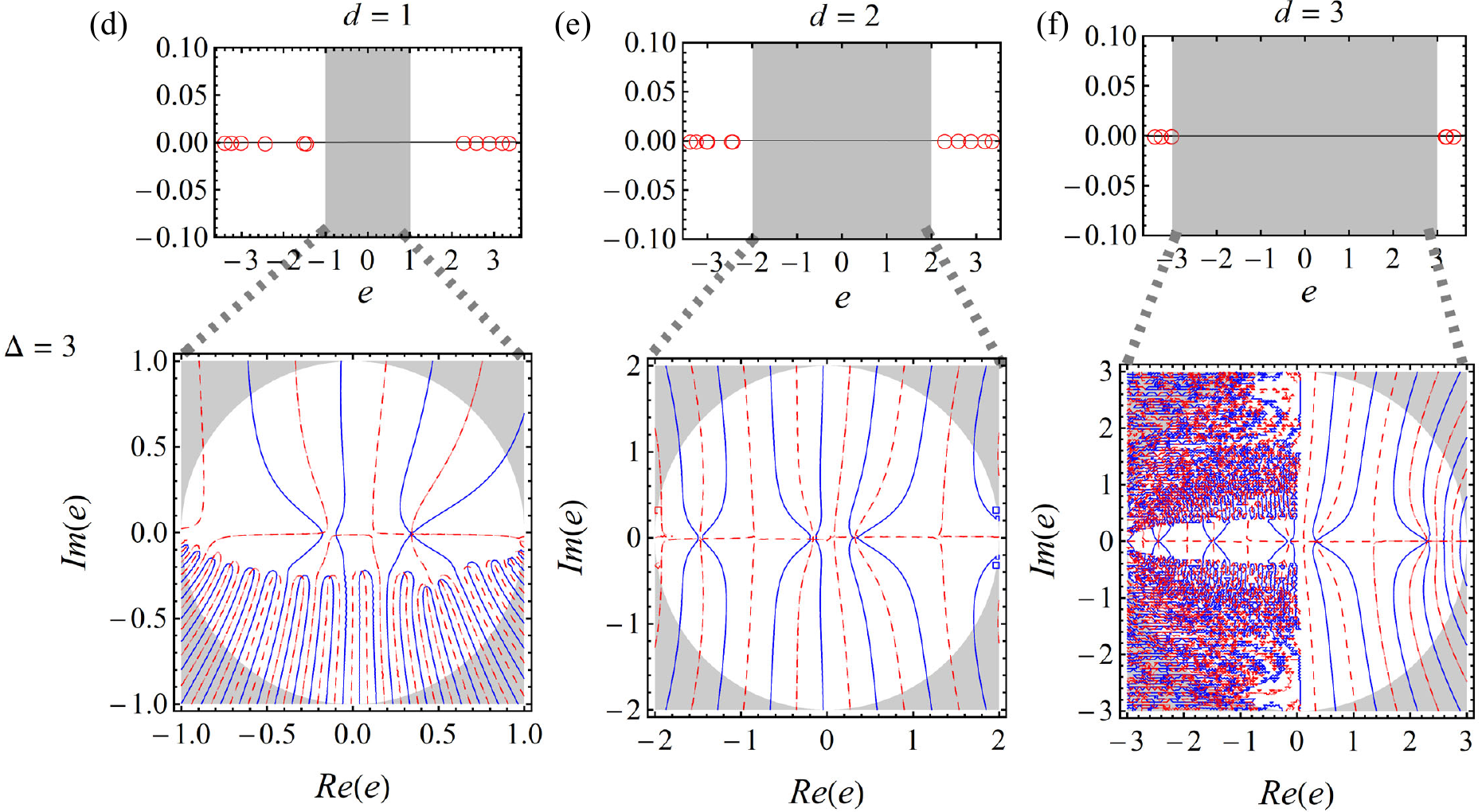}
\caption{(color online) Plots for the zero determinant of the coefficient matrix in Eq. \eqref{seq} vs. $e$ (rescaled in units of $2J$) when  $d=1, 2,$ and $3$ respectively. For definiteness,  $N_s=21$ and $\phi=-0.6\pi$ are selected.  For these plots, $\lambda \equiv 1$, and  $g=0.1$ is selected   to guarantee the validity of the rotating-wave approximation.  The panels (a), (b) and (c) investigate the case $\Delta=1$, for which the system is in the delocalized phase ($\Delta/\lambda<2$). Whereas, the panels (d), (e) and (f) investigate the case $\Delta=3$, for which the system is in the localized phase ($\Delta/\lambda>2$). In each panel, the top plot focus on the region $\left|e\right| > d$, where the real solutions (labeled by empty-red circle) can be found by solving Eq. \ref{seq}. For panel (c), since no real $e$ is found in this region, it thus is left empty.   The bottom plot focus on the region $\left|e\right| < d$. In this case, the complex solution is determined by crossing points of the solid- blue (the real part of determinant) and red- dashed (the imaginary part of determinant) line. The grey color denotes the regions outside  $\left|e\right| < d$, that is meaningless in this context.}
\label{fig:level}
\end{figure*}

To decide $e$, the determinant of the coefficient matrix of Eq. \eqref{seq} is plotted in Figs. \ref{fig:level} for different values of $\Delta$ and $d$. In all plots, $\phi=-0.6\pi$ is selected, and $\lambda=2J\equiv 1$ is assumed. To find the influence of bath on the  system, the plots focus mainly on the energy region covered by the eigenvalues of $H_S$. Thus, two situations are considered, i.e., $\left|e\right| > d$ and $\left|e\right| < d$. As illustrated in  Fig. \ref{fig:level},  the real solutions can occur when $\left|e\right| > d$. In this case, the corresponding reduce eigenstate is a bound state, which depicted the robustness of excitation  against spontaneous emission in the bath \cite{john}. Furthermore, by a thorough examination, it is found that the  reduced  eigenstate overlaps significantly with the unperturbed eigenstate of $H_S$. In this sense, the bound state can be considered the renormalization of the eigenstate in $H_S$ induced by coupling to the bath.

This situation is different for $\left|e\right| < d $.  In the bottom plot of each panel in Fig. \ref{fig:level}, a contour plot is provided for the zero determinant of the coefficient matrix of Eq. \eqref{seq}, where the blue- solid and red- dashed respectively characterize the real and imaginary part of determinant. The complex solution, $e$, can be found, which corresponds to the crossing points of blue- solid and red- dashed lines. A common feature in these plots is the occurrence of discrete zero points close  to the axis $\text{Im}(e)=0$.  By a thorough examination show in Fig.\ref{fig:a1} in Appendix, it is found for $d=1$ that the imaginary parts of these zero points have a magnitude in the order of $\sim 10^{-3}$. With a increase of $d$, the crossing points are populated more close to the axis $\text{Im}(e)=0$, as shown in Fig. \ref{fig:a1} (b), (c), (e), and (f).  Another interesting observation for $d=1$ is the appearance of  additional zero points populated in the  plane $\text{Im}(e)<0$, as shown in Figs. \ref{fig:level} (a) and (d). It is clear that these crossing points had a large imaginary parts, which characterizes a  faster decaying of excitation in the bath. This picture does not happen for $d=2$ and $3$, as shown in Figs. \ref{fig:level} (b), (c), (e) and (f). The different behavior for $d=1$ and $d=2,3$ manifests the influence of dimensionality of the bath. Additionally, it is noted for $d=3$ that a complicated behavior is observed in the region $\text{Re}(e)<0$, as shown by the bottom in Figs. \ref{fig:level} (c) and (f). Moreover, by a further examination, it can still occur up to the numerical precision of order $10^{-12}$. Unfortunately, we cannot provide an explanation for this observation.

Finally, it is should be pointed out that  the above observation is not dependent on the specific value of  $\phi$. The reason for this choice is the existence of two edge modes. Thus, the dynamics of excitation in the system would present rich properties.  In addition, the physics  of AAH model  are insensitive to the number of atomic site $N_s$ ($N_s$ being a Fibonacci number). For a large $N_s$, a numerical evaluation is computationally demanding. Thus,  $N_s=21$ is selected in this paper.

\begin{figure*}
\center
\includegraphics[width=15cm]{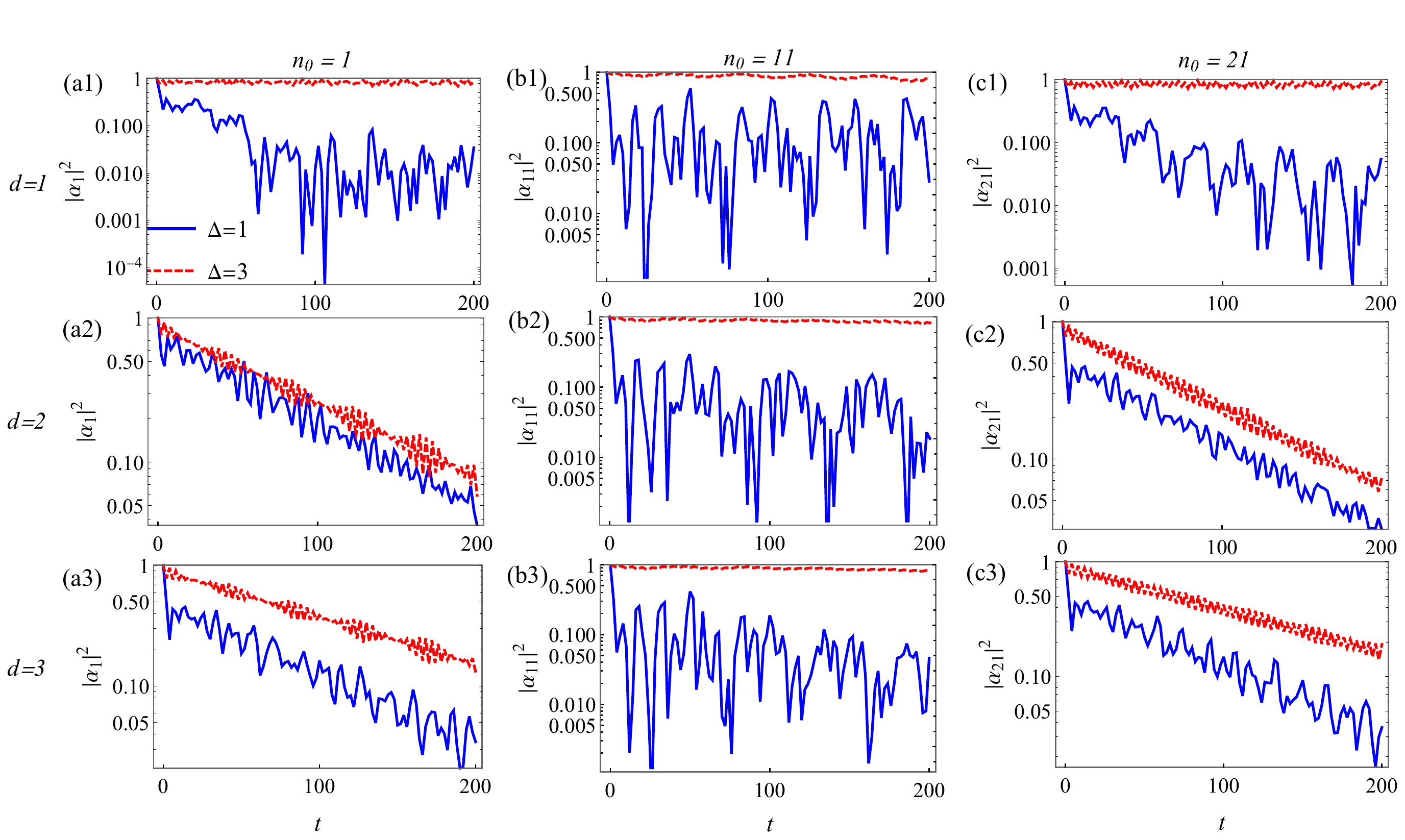}
\caption{(Color online) The logarithm  plots for the revival probability $\left|\alpha_{n_0}\right|^2$  of  single excitation initially located at atomic sites  $n_0=1, 11$ and $21$, when  $\Delta=1$ (blue- solid line) and   $\Delta=3$ (red- dashed line). $t$ is scaled in units of $(2J)^{-1}$.  The other  parameters are selected as the same as those shown   in Fig. \ref{fig:level}}
\label{fig:systemevo}
\end{figure*}

\section{The time  evolution of population }

In this section, the time evolution of  population of excitation is investigated in both atomic chains and  the lattice bath.  By the time-dependent Schr\"{o}dinger equation,  $\alpha_n(t)$ can be decided by the following equation $(\hbar \equiv 1)$
\be\label{evolution}
\mathbbm{i}\frac{\partial }{\partial t}\alpha_n(t)= \alpha_{n+1}(t) + \alpha_{n-1}(t)+ \Delta \cos(2\pi \beta n +\phi)\alpha_n(t) &\nonumber\\
- \mathbbm{i} g^2\sum_{m=1}^{N_s} \mathbbm{i}^{d \left|n-m\right|}\int_0^t \text{d}s \alpha_m(s)\left[J_{\left|n-m\right|}\left(t-s\right)\right]^d, &
\ee
where $\mathbbm{i}=\sqrt{-1}$,  and $J_{\left|n-m\right|}\left(t-s\right)$ is the Bessel function of the first kind. To obtain Eq. \eqref{evolution}, it is assumed that  the excitation is initially located in the atomic chains. Due to the existence of memory kernel  $\left[J_{\left|n-m\right|}\left(t-s\right)\right]^d$,  $\alpha_n(t)$ is correlated to its past value. To decide $\alpha_n(t)$, we first transform the integral into the summation with a suitable step length. By solving Eq.  \eqref{evolution} iteratively, $\alpha_n(t)$ can be determined for any time $t$. However, the iteration  becomes computationally expensive over an extended evolution period and a large size system.  Accordingly, the following evaluation is restrict to a  evolution time $t=200$  and $N_s=21$.

The population of excitation in the momentum space of the bath can be determined by the amplitude
\be\label{betak}
\beta_{\mathbf{k}}(t)= - \mathbbm{i} \frac{g}{\left(2\pi \right)^{d/2}} \int_0^t \text{d}s e^{\mathbbm{i}(t-s) \omega(k)}\sum_{n=1}^{N_s} e^{- \mathbbm{i}\mathbf{k} \cdot \mathbf{r}_n}\alpha_n(s).
\ee
Using Fourier transformation, the amplitude of population in the position space of the bath can be written as
\be\label{betar}
\beta_{\mathbf{r}}(t)&=&\frac{1}{\left(2\pi\right)^{d/2}}\int_{-\pi}^{\pi}\text{d}k_1 \int_{-\pi}^{\pi}\text{d}k_2\cdots \int_{-\pi}^{\pi}\text{d}k_d \beta_k(t) e^{ \mathbbm{i}\mathbf{k} \cdot \mathbf{r}}\nonumber\\
&=& - \mathbbm{i} g \int_0^t \text{d}s \sum_{n=1}^{N_s} \alpha_n(s) \otimes_{j=1}^d  \mathbbm{i}^{\left|n +r_j\right|} J_{\left|n+r_j\right|}\left(t-s\right)
\ee
where $\mathbf{r}=\left(r_1, r_2, \cdots, r_d\right)$ is a vector in the $d$-dimensional real space. Evidently,  both $\beta_{\mathbf{k}}(t)$ and $\beta_{\mathbf{r}}(t)$ are directly correlated to  past states of the system. Consequently, the spreading of excitation in the bath is correlated to the state of system.

\begin{figure}
\center
\includegraphics[width=9cm]{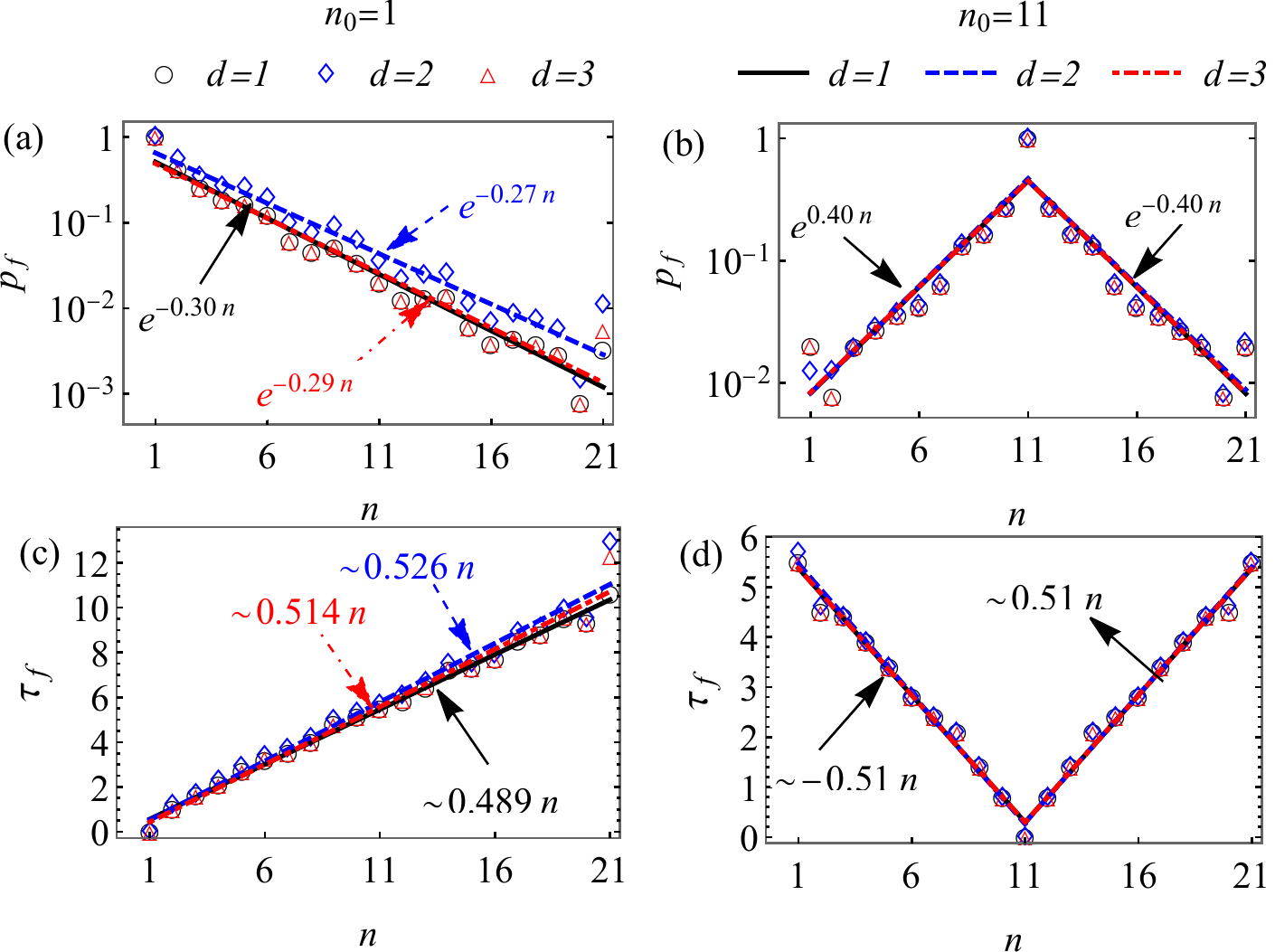}
\caption{(Color online) Plots for the first peak $p_{f}$ for the population of excitation on atomic site $n$ and its occurring  time  $\tau_{f}$ (in unit of $2J$). For these plots, $\Delta=1$ was selected. The other parameters are the same as those shown in Fig. \ref{fig:level}. In panels (a) and (c), the fitting parameters are presented by the colors according to  the value of $d$. However, in panels (b) and (d), the fitting parameter for different values of $d$ is nearly the same. Thus, only one parameter is presented.   }
\label{fig:shortimeS}
\end{figure}

\begin{figure}
\center
\includegraphics[width=9cm]{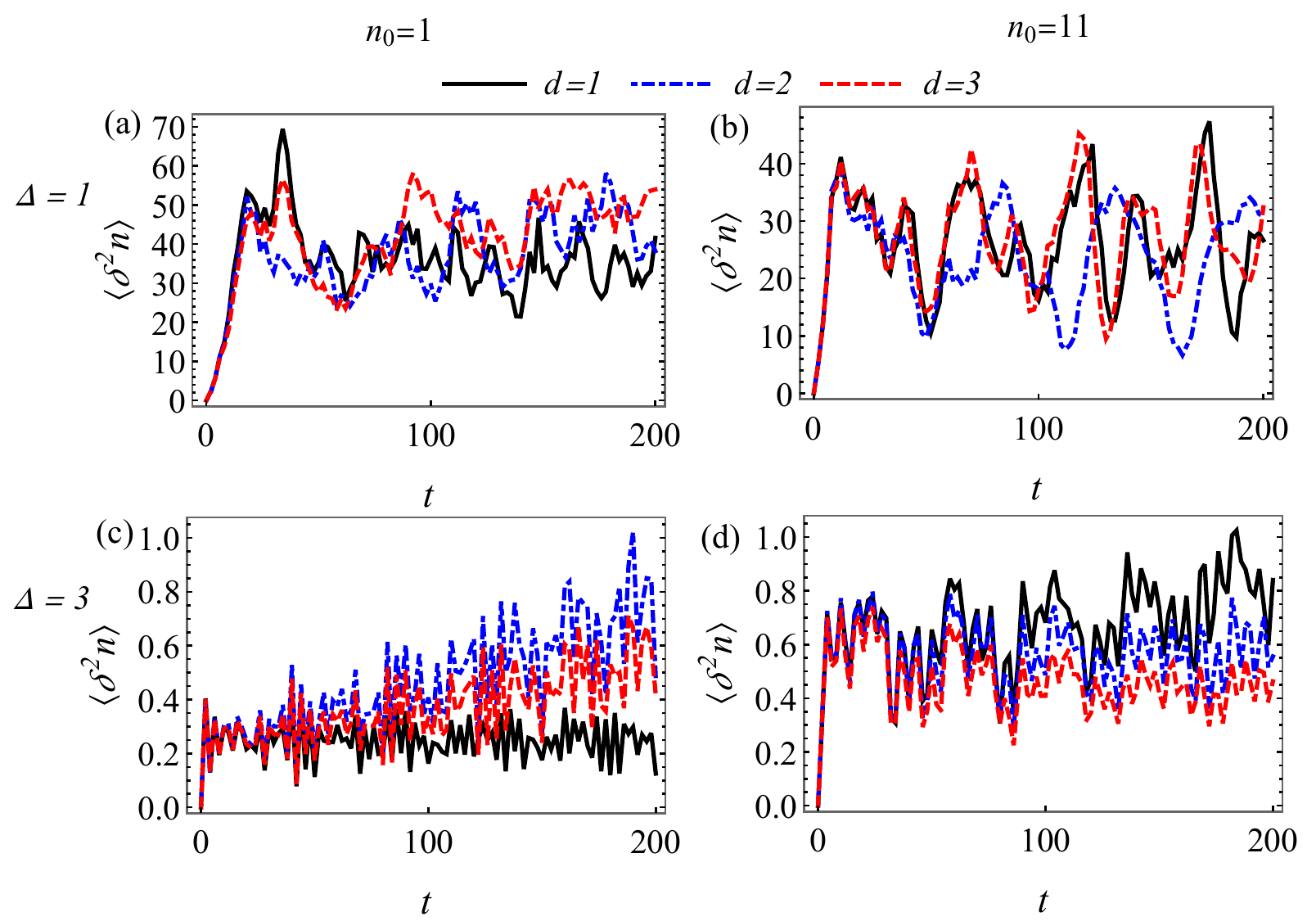}
\caption{(Color online) Comparative plot of $\langle\delta^2 n\rangle$ for different values of $d$, $\Delta$ and $n_0$. The remaining parameters are the same as those shown in Fig. \ref{fig:level}.}
\label{fig:fluctuationS}
\end{figure}

\subsection{Population evolution in the system}

In this subsection, the spreading of excitation in the atomic chain is studied by evaluating the revival probability $\left|\alpha_{n_0}\right|^2=\left|{_{n_0}\bra{1}}e^{-\mathbbm{i}H t}\ket{1}_{n_0}\right|^2$. In addition, it is supposed that the excitation is initially located at the ends (labeled by $n_0=1$ and $21$) or in the middle (labeled by $n_0=11$)  of atomic chain.  The choice for initial state is related to the fact that AAH model can have two in-gap edge states, for which the excitation becomes highly localized at the two ends of chain. Thus, the initial states for $n_0=1$ and $21$ correlated strongly  to the two edge states. While, the initial state for $n_0=11$ can be considered as a superposition of the band states of the system. So, it is expected that dependent on the initial state, the evolution of revival probability, as well as the spreading of excitation in the system,  can show distinct behavior.

Let first focus on the time  evolution of excitation initially at $n_0=1$ and $21$.  It is evident that related to $d$ or $\Delta$,  the revival probability can present different behaviors , as shown in Fig. \ref{fig:systemevo} (a1)-(a3) and (c1)-(c3). For $d=1$ and $\Delta=1$, both $\left|\alpha_{1}\right|^2$  and $\left|\alpha_{21}\right|^2$  decay rapidly at first, and then  become stable, as shown by the solid-blue lines in Fig. \ref{fig:systemevo}(a1) and (c1). In contrast, with a increase of $\Delta$,  $\left|\alpha_{1}\right|^2$  and $\left|\alpha_{21}\right|^2$ display robustness against decaying. This observation can be understood by the fact  that  the initial states of $n_0=1$ and $21$ overlap significantly with the in-gap edge states in AAH model. For $\Delta=1$, the edge levels are embedded into the continuum of bath. As a result, the excitation is emitted spontaneously into the bath. However, for $\Delta=3$, the edge states are renormalized as bound states since the corresponding levels lie outside the continuum of bath. Thus, because of the energy gap between the edge levels and the spectrum of bath, the bound state is robust against dissipation. In Appendix, the plots  for the  inverse participation ratio (IPR), defined as $\text{IPR}=\sum_{n=1}^N \left|\alpha_n(t)\right|^4$, is presented to characterize the spreading of excitation in the atomic chain. IPR characterizes the property of excitation localized in the system. As shown in Fig. \ref{fig:ipr} (a1) and (c1), IPR becomes steady around  $\sim 0.01$ when $\Delta=1$. This observation means that although coupled to the bath, the excitation can be retained in the system by a finite probability.  In contrast, when $\Delta=3$, IPR has a value close to  1, It means that the excitation can be retained on its initial position.

However, the picture is different for a increase of $d$. As shown in Figs. \ref{fig:systemevo} (a2), (a3), (c2) and (c3), both $\left|\alpha_{1}\right|^2$  and $\left|\alpha_{21}\right|^2$  show an exponential decay whenever $\Delta=1$ or $3$.  This difference is a result of the observation in Fig.\ref{fig:level}, in which the levels $e$ shows significantly correlation to the dimensionality of bath. For $d=1$, besides of the solutions close to the axis $\text{Im}(e)=0$,  there are  the additional complex solutions as shown in Figs. \ref{fig:level} (a) and (d). These complex levels have a large imaginary part, and then accelerate the decaying of excitation in the bath. Furthermore, the interference in the complex levels also makes the revival probability stable rapidly shown Fig. \ref{fig:systemevo}(a1) and (c1). In constrast, for $d=2$ and $3$, the reduced energy level can only be found close to the axis $\text{Im}(e)=0$. As a consequence, the decay of excitation is dominated uniquely by the reduced energy state significantly overlapped with the initial state. This is reason for the exponential decaying of revival probability, observed in  Figs. \ref{fig:systemevo} (a2), (a3), (c2) and (c3). Similar observation can be found for IPR shown in Fig. \ref{fig:ipr} (a2), (a3), (c2) and (c3).

As for $n_0=11$,  the revival probability seems insensitive to  the dimension of bath. In Fig. \ref{fig:systemevo} (b1)- (b3) and Fig. \ref{fig:ipr} (b1)- (b3), $\left|\alpha_{11}\right|^2$ and  the corresponding IPR are presented respectively. A subtle point is for the case $\Delta=1$ that  $\left|\alpha_{11}\right|^2$ rapidly  becomes  stable when $d=1$, while descents slowly when $d=2$ or $3$. This observation can also be attributed to the fact the initial state of $n_0=11$ is a superposition of band states of $H_S$: The interference in the band states would make the decaying slowly. Moreover, the occurrence of additional solutions for $d=1$ can speed up the decaying to a steady value. In contrast, for $\Delta=3$, $\left|\alpha_{11}\right|^2$ shows a small descent,  independent of the value of $d$. This picture can be attributed to the strong localization in the system, which prevents the decaying of excitation into the bath.

\subsection{Spreading of excitation in the atomic sites}

To gain further information of the evolution of excitation, the spreading of excitation in the system was explicitly studied for $\Delta=1$ by evaluating  the first peak $p_{f}$ for the population  of excitation on the atomic sites ( Figs. \ref{fig:shortimeS} (a) and (b)) and its occurring time $\tau_{f}$( Figs. \ref{fig:shortimeS} (c) and (d)).  Actually $p_{f}$ characterizes the property of dissipation, and $\tau_f$ provides the information of  spreading velocity of excitation in the system \cite{quantumwalks}.  By linear fitting of $\tau_f$, one find that the propagation velocity is $\sim 2\text{sites}/(2J)^{-1}$, independent of $d$.  However,  $p_{f}$ decayed exponentially with the distance $\left|n-n_0\right|$ that means a dissipative propagation of excitation in the system.  Also,  $p_{f}$  displays very weak dependence on the value of $d$. This observation shows that at early stage of evolution, the propagation of excitation in the system  would be irrelevant to the property of the environment \cite{beau}. Additionally,  it is noted that the fitting parameters for both $p_f$ and $\tau_f$ show evident correlation to the initial state as shown in Figs. \ref{fig:shortimeS}. This picture means that the initial condition would affect the spreading of excitation in the system.  The situation for $n_0=21$ is similar as $n_0=1$, which is not presented.

The spreading of excitation becomes different over an extended time. In Fig. \ref{fig:fluctuationS}, the variance of the position of excitation within the atomic site, i.e.,  $\langle\delta^2 n\rangle= \frac{\sum_n\left|\alpha_n\right|^2\left(n-\langle n\rangle\right)^2}{\sum_n \left|\alpha_n\right|^2}$, is plotted. For $n_0=1$, it is evident that,  as shown in Figs.\ref{fig:fluctuationS} (a) and (c),   $\langle\delta^2 n\rangle$ rapidly tends to be stable when $d=1$ wether $\Delta=1$ or $3$. However, by a increase of $d$, $\langle\delta^2 n\rangle$ displays a weak tendency of ascent when $\Delta=1$, while this tendency becomes more significant when $\Delta=3$. This picture implies that the transport of excitation may be changed by  bath when the system is highly localized.  In contrast, for $n_0=11$, a stable oscillation of  $\langle\delta^2 n\rangle$ can be observed for all   values of $d$ when $\Delta=1$, as shown in Fig.\ref{fig:fluctuationS} (b). Similar picture can be found for $\Delta=3$ shown in Fig.\ref{fig:fluctuationS} (d), except of the case $d=1$, for which a weak ascent of $\langle\delta^2 n\rangle$ is noted. This picture can be understood by the superposition  property of the initial state  $n_{0}=11$, which makes the system reach stability quickly.

These observations can be understood by the property of localization in the system. When $\Delta=1$, the system is in the extended phase, in which excitation dynamically tends to populate on the atomic site with equal probability. In this case, the spreading of excitation in the atomic sites is dominated by the property of the system, independent of dimensionality of bath. In contrast, when $\Delta=3$, the system is in the localized phase, in which excitation is inclined to retain its initial position. This is the reason that the amplitude of the variance is significantly smaller than that for $\Delta=1$. The different behaviors for $n_0=1$ and $n_0=11$  shown in Figs. \ref{fig:fluctuationS} (c) and (d), is an exemplification of the distinct property of the initial state.

Finally, it should be pointed out that the fluctuation of $\langle\delta^2 n\rangle$, as well as revival probability, is a result of the backaction effect of lattice bath. Phenomenally, one can understand this effect as the following process. By the local interaction $H_I$, the excitation is transferred to a lattice site in the bath. Because of the periodic boundary condition of lattice bath, the excitation can travel throughout the entire  bath, and in a later time may return to its initial position in the bath with a finite probability. Consequently, also by $H_I$, the excitation can be transferred back to the system with a finite probability. In this process, the exchange of excitation between the system and bath gives rise to the fluctuation of $\langle\delta^2 n\rangle$ and revival probability.
In this point, the spreading of excitation in the bath would become significantly relevant to the population of excitation in the system. Thus it is very interesting to find the characters of evolution of excitation in the bath.

\begin{figure}
\center
\includegraphics[width=8.5cm]{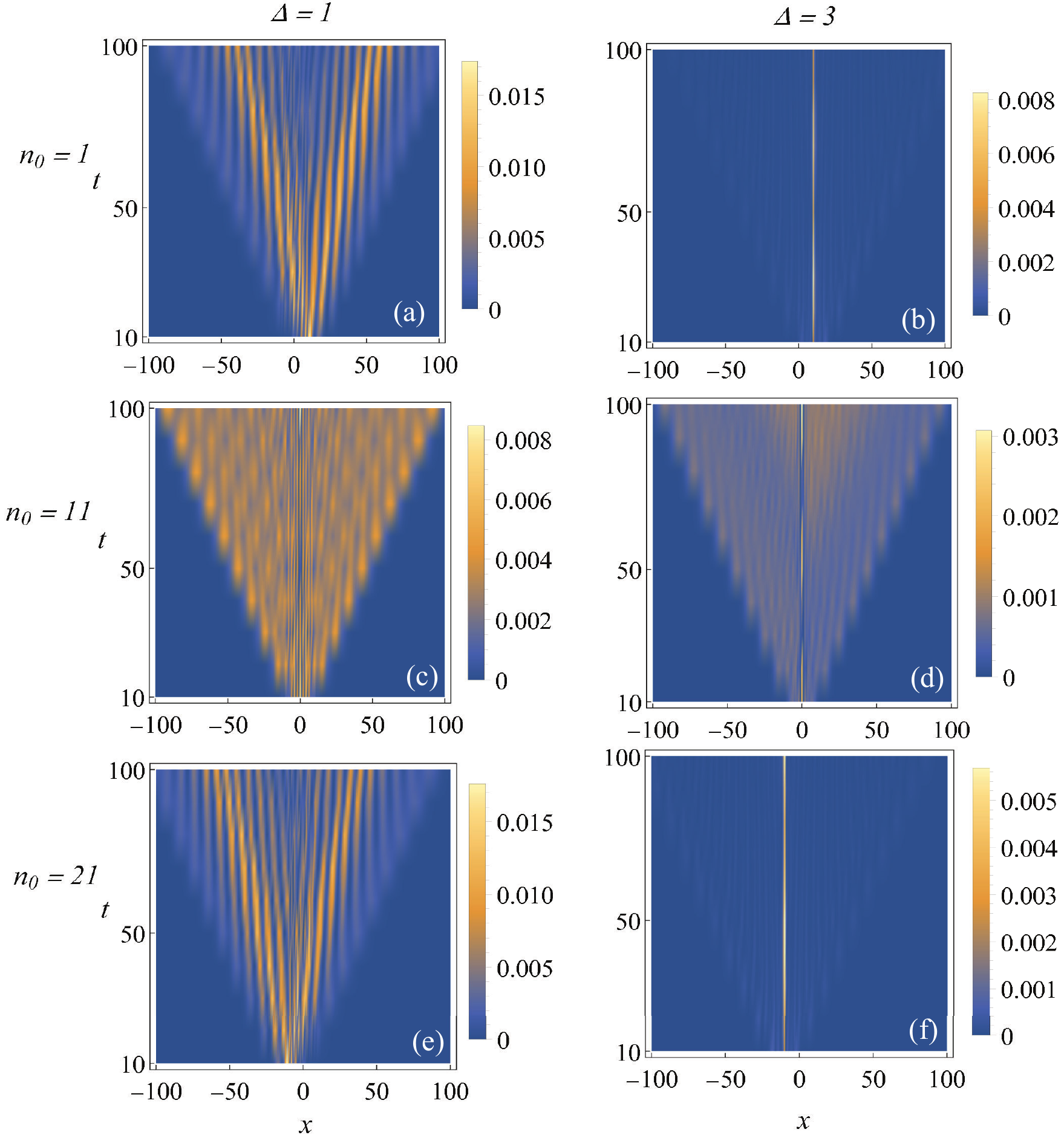}
\caption{(Color online) The density plots for the time evolution of $\left| \beta_{\mathbf{r}}(t)\right|^2$  when the  lattice bath  is one-dimensional ($d=1$). $t$ is scaled in units of $(2J)^{-1}$.  $x$ labels the site of lattices in the bath, and the origin point ($x=0$) is placed  at the lattice site coupled to the center atomic site $n=11$. The atomic chain is organized along the lattice bath.  For these plots, the number of lattices in the bath $N_b=201$ is chosen.  The remaining parameters are the same as those in Fig. \ref{fig:level}.}
\label{fig:1Dconfiger}
\end{figure}

\begin{figure*}
\center
\includegraphics[width=16cm]{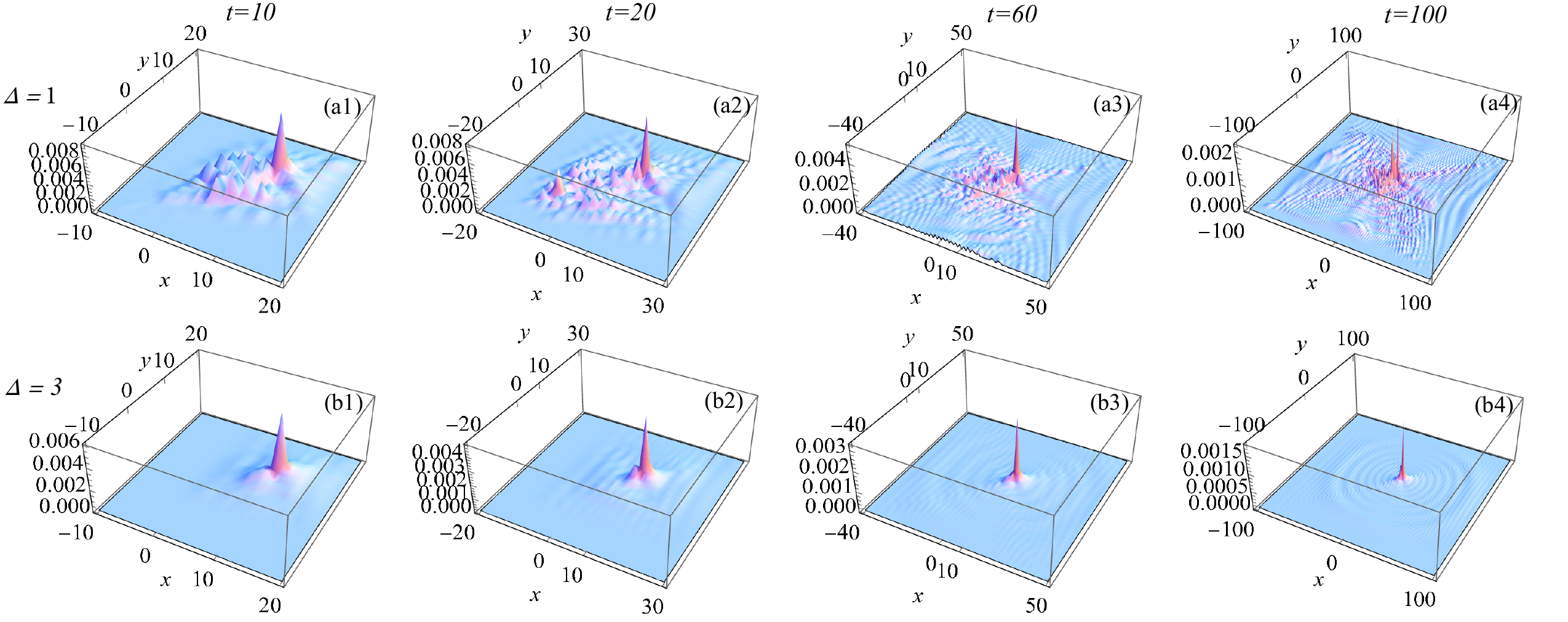}
\includegraphics[width=16cm]{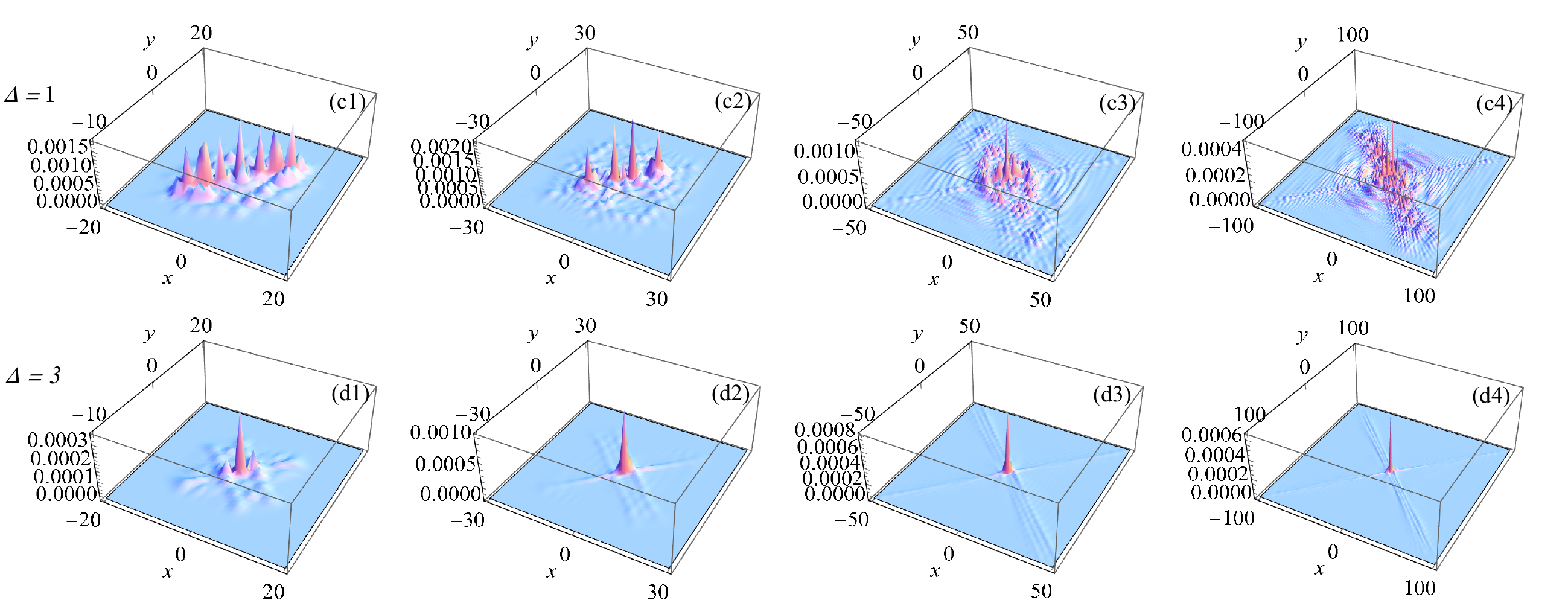}
\caption{(Color online)  Plots of  $ \left| \beta_{\mathbf{r}}\right|^2$ for different evolution times when the  bath is a square lattices with dimension $201\times 201$. $(x, y)$  labels the coordinates of  the lattice sites in the bath. The atomic chain is placed along the diagonal direction of the square lattices, and the origin point $(0, 0)$ is placed at the lattice site coupled to the atomic site $n=11$.  Thus, for the atomic sites $n=1$ or $21$, the coordinates  of corresponding coupled lattice sites are $(10, 10)$ or $(-10, -10)$. The other parameters are same to those in Fig. \ref{fig:level}. To more clearly illustrate the distribution of excitation in the lattice bath, the plots focus on the region where $ \left| \beta_{\mathbf{r}}\right|^2$ is non zero. Panels (a1)-(a4) and (b1)-(b4) are plotted for $n_0=1$; Panels (c1)-(c4) and (d1)-(d4) are plotted for $n_0=11$.}
\label{fig:2Dconfiger}
\end{figure*}

\begin{figure*}
\center
\includegraphics[width=16cm]{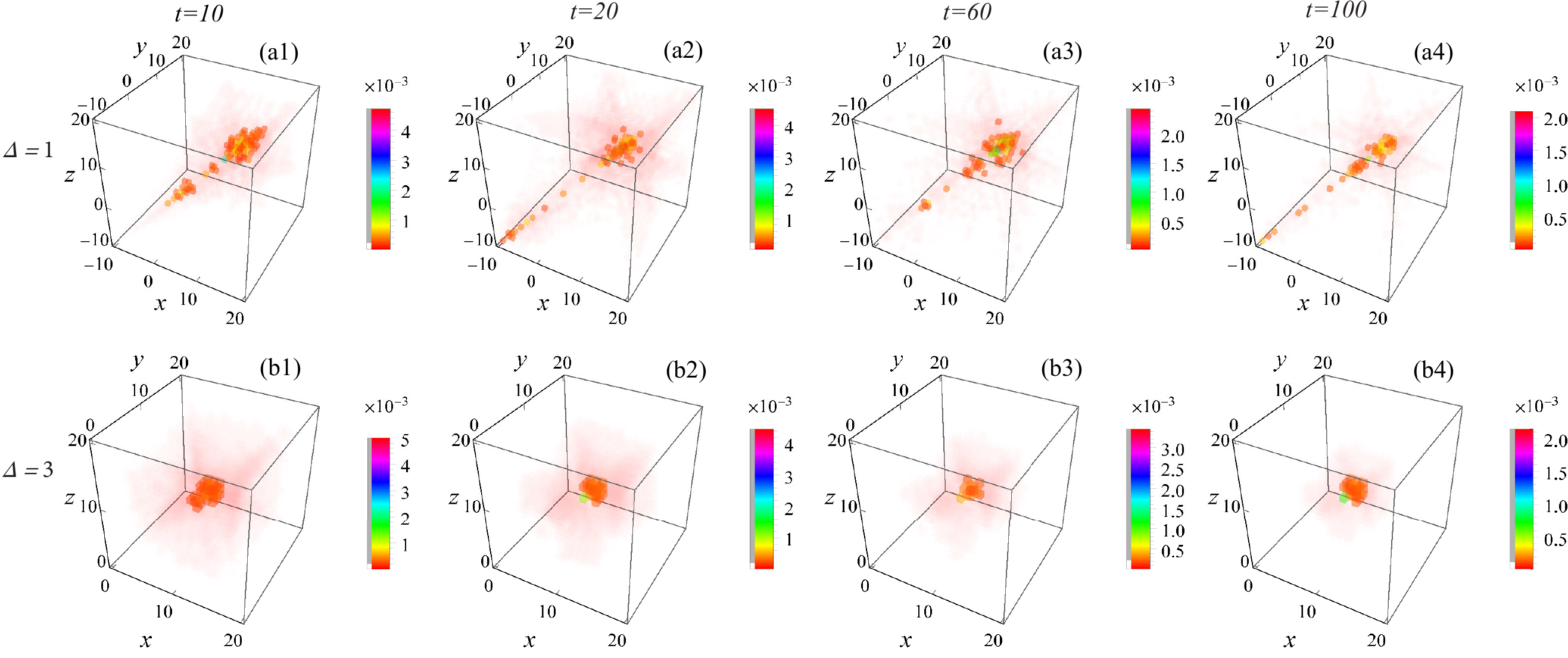}
\includegraphics[width=16cm]{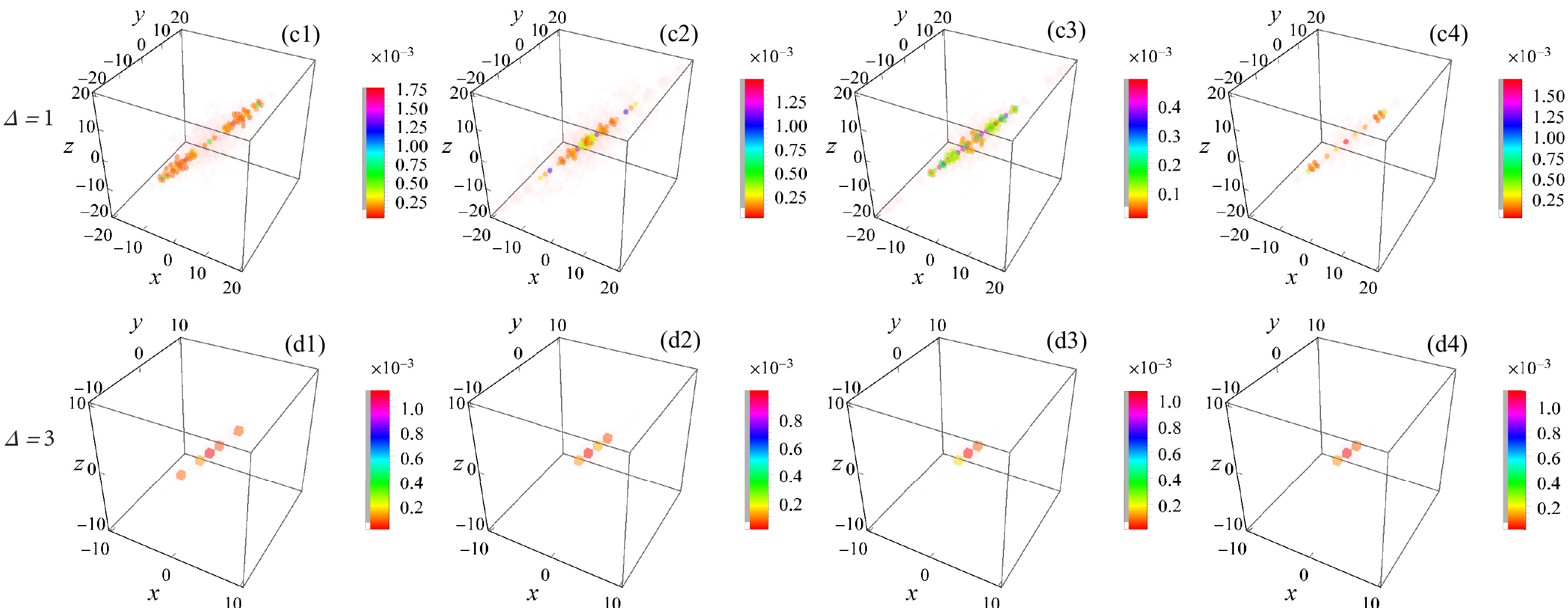}
\caption{(Color online)The three-dimensional density plots of  $\left| \beta_{\mathbf{r}}\right|^2$ for different evolution times when the bath is a cubic lattices with dimension  $51\times 51\times 51$.  $(x, y, z)$ denote the  coordinates of cubic lattice sits in the bath. The atomic chain is placed along the  body-diagonal direction of cubic lattice, and the original point is placed on the lattice site coupled to the atomic site $n=11$. Thus, for atomic sites $n=1$ or $21$, the corresponding coupled lattice site is at $(10, 10, 10)$ or $(-10, -10, -10)$. The remaining parameters are the same as those in Fig. \ref{fig:level}. In addition, for a clear demonstration, the plots focus mainly on the region where $ \left| \beta_{\mathbf{r}}\right|^2$ is finite. Panels (a1)-(a4) and (b1)-(b4) are plotted for $n_0=1$; Panels (c1)-(c4) and (d1)-(d4) are plotted for $n_0=11$.}
\label{fig:3Dconfiger}
\end{figure*}

\subsection{Population evolution in the bath}

Regarding the finiteness of the bath, it is interesting to find how excitation spreads in the bath. As shown in Eq. \eqref{betar}, the propagation of excitation in the bath is significantly correlated to the past states of the system. Hence,  it is anticipated that the dynamics of excitation in the bath could be used to extract useful information of the system. Moreover,  finding the distinct dynamics of excitation in the bath may provide additional insight of the excitation dynamics in the system. For this purpose,  the time evolution of the  population  $ \left| \beta_{\mathbf{r}} \right|^2$ is studied  for $d=1, 2$, and $3$ respectively in this subsection.

\emph{${d=1}$}  The time evolution of  $ \left| \beta_{\mathbf{r}} \right|^2$ is plotted in Figs. \ref{fig:1Dconfiger} For these plots, the atomic chain is placed along the lattice direction, and $x=0$ is set on the lattice site coupled to the center atomic site. To avoid the boundary effect, the time evolution is restricted up to $t=100$ in this case. Evidently, depending on the initial states, the propagation of excitation in the bath can display two different features.  Let first focus on the excitation initially located at $n_0=1$ or $21$, for which the locally coupled lattice sites in bath is at $x=11$ or  $-11$. For $\Delta=1$,  the excitation propagates in a wavepacket manner along  two opposite directions as shown in Figs. \ref{fig:1Dconfiger} (a) and (e). In contrast, when $\Delta=3$,  the spreading is  compressed completely   as shown in Fig. \ref{fig:1Dconfiger} (b) and (f). Thus, the excitation appears to populate only  on the lattice site $x=11$ or $-11$. This picture comes from the fact that the initial state is  renormalized as a bound state, and thus the atomic site and coupled lattice site become entangled strongly. As a result, excitation can only be populated oscillatorily  between the two sites. The situation is different for $n_0=11$.   As shown in Fig. \ref{fig:1Dconfiger}(c) and (d), the excitation tends to be populated on all lattice sites of bath when $\Delta=1$. While,  the spreading  is significantly compressed when $\Delta=3$   as shown in Fig. \ref{fig:1Dconfiger}(d). This picture is a result of the localization in the system, which precludes the excitation from decaying into bath.

The differen behaviors of the population evolution between  $n_0=11$ and $n_0=1$ or $21$ when $\Delta=1$ can be attributed to the effect of initial state. For $n_0=1$ or $21$, the initial state is strongly correlated to the in-gap edge state of the system.  Thus, the population of excitation in the atomic sites may behave localized. As a result, since $ \left| \beta_{\mathbf{r}} \right|^2$ is correlated to the state of system as shown by Eq. \eqref{betar}, it is not surprise that the propagation of excitation in bath may also display the character of localization. In contrast, the energy gap is absent for  the initial state of $n_0=11$. Subsequently, the coupling of system to a common both induces the hopping of excitation in the atomic sites. Similarly, by Eq. \eqref{betar},  the spreading of excitation in the bath is a result of the extensity of the state of system.

\emph{${d=2}$}  The propagation of excitation  became more complex when the lattice bath is  two- dimensional.  In Fig. \ref{fig:2Dconfiger},  $\left| \beta_{\mathbf{r}} \right|^2$ are presented for several evolution times. It is evident that the spreading of excitation in the bath displays the intimate connection to the initial state.  For $n_0=1$, a common feature is the maximal population of excitation at lattice site $(10, 10)$, which is  locally coupled  to the atomic site $n=1$. This picture is differen from the observation in Fig. \ref{fig:1Dconfiger} (a), where $\left| \beta_{\mathbf{r}} \right|^2$ can show the maximum away from the lattice site coupled to the  atomic site initially populated by excitation. This can be understood by the fact that the increase of dimension may provide addition decaying channels of excitation. Thus,  the excitation in the bath can spread along different directions.   This point becomes apparent by the time evolution of  $\left| \beta_{\mathbf{r}} \right|^2$ shown in Fig. \ref{fig:2Dconfiger} (a1)- (a4): the excitation in the bath spreads  mainly along the diagonal and off-diagonal directions. However, with a increase of $\Delta$, the atomic site $n_0=1$ and the lattice site $(10, 10)$ is strongly entangled because of the occurrence of bound state. Thus, the excitation spreads in the bath much like a single particle. This feature is illustrated in Fig. \ref{fig:2Dconfiger} (b1)- (b4). It is found that the excitation can spread isotropically in the bath. Similar observation can be found for $n_0=21$, which is not presented here.

For $n_0=11$  shown in Fig. \ref{fig:2Dconfiger} (c1)- (c4), $\left| \beta_{\mathbf{r}} \right|^2$ can be pronounced on the lattice sites belong to the line segment between  $(-10, 10)$ and $(10, 10)$, where the lattice sites are coupled to the atomic sites. This picture is a result of the hopping of excitation in the atomic sites induced by coupling to the bath. Thus, because of the local interaction $H_{int}$, the excitation can be transferred between the atomic and lattice sites. Moreover, outside this special region, the spreading of excitation in the bath can also be found along the diagonal and off-diagonal directions. With a increase of $\Delta=3$, $\left| \beta_{\mathbf{r}} \right|^2$ becomes pronounce significantly only at site $(0, 0)$. The spreading of excitation in diagonal and off-diagonal directions of the bath is compressed greatly. The reason is that the strong localization in the system precludes the excitation from hopping in the atomic sites. Consequently, it is difficult to transfer the excitation from the system to the bath through the interaction $H_{int}$.

\begin{figure}
\center
\includegraphics[width=8.5cm]{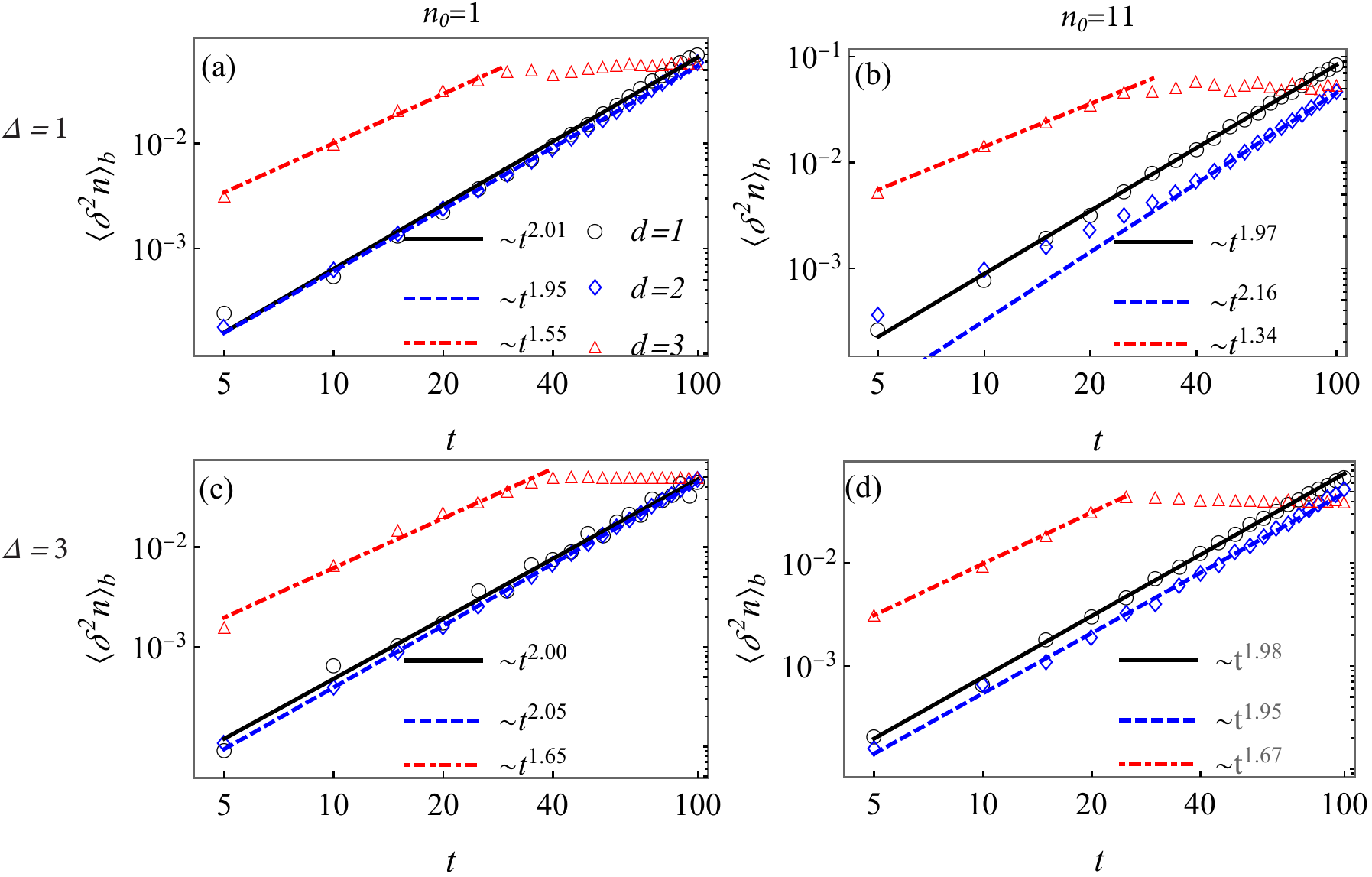}
\caption{(Color online) The log-log plots of $\langle\delta^2 x\rangle_b$ for different values of $d$, $\Delta$ and $n_0$. For these plots,  the number of lattice sites in the environment is selected as $201$ for $d=1$,  $201\times 201$ for $d=2$ and $51\times 51 \times 51$ for $d=3$. The remaining parameters are the same as those in Fig. \ref{fig:level}. The solid-black, dashed-blue and dot-dashed-red lines correspond to the  fitting lines for $d=1$, $2$ and $3$.  }
\label{fig:fluctuationB}
\end{figure}

\emph{${d=3}$}   In Figs. \ref{fig:3Dconfiger}, the three-dimensional density plot for $\left| \beta_{\mathbf{r}} \right|^2$ is presented. Similar to the observation for $d=2$, the distribution displays significant correlation to the initial state. For $n_0=1$, it is found  that there is a maximal value of $\left| \beta_{\mathbf{r}} \right|^2$ at lattice site $(10, 10, 10)$, where is locally coupled to the atomic site initially occupied by excitation. However, on the other lattice sites, the value of $\left| \beta_{\mathbf{r}} \right|^2$  can be finite when $\Delta=1$ as shown in Figs. \ref{fig:3Dconfiger} (a1)- (a4). While, with a increase of $\Delta$, $\left| \beta_{\mathbf{r}} \right|^2$ at lattice site $(10, 10, 10)$ becomes more pronounced, and on the other sites, $\left| \beta_{\mathbf{r}} \right|^2$ tends to be zero as shown in Figs. \ref{fig:3Dconfiger} (b1)- (b4). In Fig. \ref{fig:a3} (a)- (d) in Appendix, a plot for $\left| \beta_{\mathbf{r}} \right|^2$ on the body diagonal direction is provided. Clearly, the spreading of excitation in the bath is restricted mainly  in the region  between $x=y=z=-10$ and $10$, where the lattice sites are coupled to the atomic sites. This observation implies that the increase of dimension of bath would reduce the spreading of excitation. Similar observations can also be found for $n_0=21$, which is not presented here.

For $n_0=11$, $\left| \beta_{\mathbf{r}} \right|^2$ can be pronounced on the lattice sites away from $(0, 0, 0)$ when $\Delta=1$ as shown in Figs. \ref{fig:3Dconfiger} (c1)- (c4). Also, this picture is a manifestation of the hopping of excitation  induced by the system's coupling to the bath. However, for $\Delta=3$, $\left| \beta_{\mathbf{r}} \right|^2$ becomes much pronounce on site  $(0, 0, 0)$, and the spreading of excitation on the other sites is significantly compressed. A detail plot for  $\left| \beta_{\mathbf{r}} \right|^2$ on the body diagonal sites can be found in Figs. \ref{fig:a3} (e)- (h).

To understand the propagation of excitation in the bath, the variance of position, defined as
\be
\langle \delta^2 x \rangle_b=\frac{1}{N^2_{b}} \frac{\sum_{\mathbf{r}} \left(x -\langle x \rangle\right)^2 \left|\beta_{\mathbf{r}}\right|^2}{\sum_{\mathbf{r}} \left|\beta_{\mathbf{r}}\right|^2}.
\ee
is studied in Fig. \ref{fig:fluctuationB}, in which $\mathbf{r}=(x, y, z)$ and $N_b$ denotes the number of lattice site in the $x$ direction. With respect to the translational invariance of $H_b$, $\langle \delta^2 x \rangle_b$ is evaluated only on the $x-$direction when $d=2$ and $3$. The fluctuation can scale as $\langle \delta^2 x \rangle_b \sim t^{\nu}$, where $\nu$ is the diffusive exponent which is determined by the transport process. In ballistic transport, $\nu=2$; in diffusive transport, $\nu=1$. $1<\nu<2$ corresponds to super-diffusive transprot. Accordingly, power fitting was applied when $d=1$ and $2$ as shown in Fig.  \ref{fig:fluctuationB}. It is found that for $d=1$ and $2$,  $\nu$ had a value close to 2, which means a ballistic-like transporting of excitation in the lattice bath.  As for $d=3$, $\langle\delta^2 x\rangle_b$ becomes stable rapidly after $t \sim 30$ because of the  small size of bath. Thus, the power fitting is implemented only for data up to $t \sim 30$. It is evident as shown in Fig.  \ref{fig:fluctuationB} that the value of $\nu$ is less than $2$, which means a super-diffusive transport of excitation.

\section{conclusion}

In conclusion, the dynamics  of excitation is  studied in the Aubry-Andr\'{e}-Harper model coupled to a simple high-dimensional lattices bath. Depending on  the initial state, localization in the system, and the dimensionality of the lattices bath, the propagation of excitation in both the system and the bath can illustrate different features.

As shown in Fig. \ref{fig:systemevo}, the revival probability $\left|\alpha_{1}\right|^2$ and $\left|\alpha_{21}\right|^2$ decays super-exponentially for $d=1$ when the system is in extended phase ($\Delta=1$). In contrast, both of them decay exponentially for $d=2$ and $3$. This observation is a result of the distribution of the reduced energy levels  determined by Eq. \eqref{seq}. As shown in  Figs. \ref{fig:level}, depending on  the dimensionality of the bath, the reduced energy levels may display two different characteristics.  For $d=1$, besides of  the energy levels close to the axis $\text{Im}(e)=0$,   the additional solutions with imaginary parts having a magnitude of order $10^{-3}$ can also be found. The interference between the two types solutions induces the fast decaying  and rapid stability of  $\left|\alpha_{1}\right|^2$ and $\left|\alpha_{21}\right|^2$. However, for $d=2$ and $3$, only energy levels close to the imaginary axis can be found, which can be considered as the  renormalizatio of level in the system. Thus,  the imaginary part of the complex levels gives rise to the exponential decaying of revival probability. For a increase of $\Delta$, the bound states can occur,  which is responsible for the robustness of revival probability.

The situation is different for the initial state of $n_0=11$. As shown in Fig. \ref{fig:systemevo},  $\left|\alpha_{11}\right|^2$ decays more quickly. This difference comes from the distinct property of initial state. For $n_0=1$ or $21$, the corresponding initial state overlaps significantly with the in-gap edge state in the system. Because of the occurrence of energy gap, the hopping of excitation in the system, induced by coupling to the bath, is compressed.  However, the energy gap is absent for $n_0=11$. The coupling to the bath makes the energy levels interfered, which enhances the decaying of excitation in the system.

The different property between $n_0=1 (21)$ and $11$ is also responsible for the distinct spreading of excitation in the bath. As shown in Section IIC, the  excitation in the bath can spread in  different manners related to the dimension of bath $d$ and the initial state.  For $n_0=1$, excitation can spread in a wavepacket manner when $d=1$. However, by a increase of $d$, the excitation can spread along different directions in the bath. The combination of local interaction $H_{int}$ and the existence of energy gap induces that the population of excitation is pronounced on the lattice site coupled to the initial atomic site of excitation. In contrast, for $n_0=11$, the hopping of excitation in the system can result in the diffusion of excitation in the bath. Finally, with a increase of $\Delta$, the system displays strong localization. Consequently, whether for $n_0=11$ or $n_0=1 (21)$, the population of excitation in the bath become sharply pronounced on the lattice site coupled to the initial atomic site of excitation. The propagation of excitation in the other lattice sites is greatly compressed.

By a thorough study of the population evolution of single excitation in both the system and a finite bath, we demonstrated   the significant influence of localization in the system and the dimensionality of the  bath, as well as the initial state.  Our study demonstrates clearly the nontrivial inter-effect between the system and its environment under the non-Markovinity. We believe that the findings in this work are useful in quantum information processing, especially in the regime of quantum control. An open question is for the influence of the particle interaction. However, the exact treatment of non-Markovian dynamics in the context of interacting many-body systems is a challenging task. This topic will be touched in a future work.

\section*{Acknowledgments}
M.Q. acknowledges the support of National Natural Science Foundation of China (NSFC) under Grant No. 11805092 and  Natural Science Foundation of Shandong Province under Grant No. ZR2018PA012. H.Z.S. acknowledges the support of National Natural Science Foundation of China (NSFC) under Grant No.11705025. X.X.Y. acknowledges the support of NSFC under Grant No. 11775048.

\renewcommand\theequation{A\arabic{equation}}
\setcounter{equation}{0}
\renewcommand\thefigure{A\arabic{figure}}
\setcounter{figure}{0}
\section*{Appendix}

In this section, some complementary statements or plots are provided to illustrate further the observation in the main text.

\begin{figure*}
\center
\includegraphics[width=15cm]{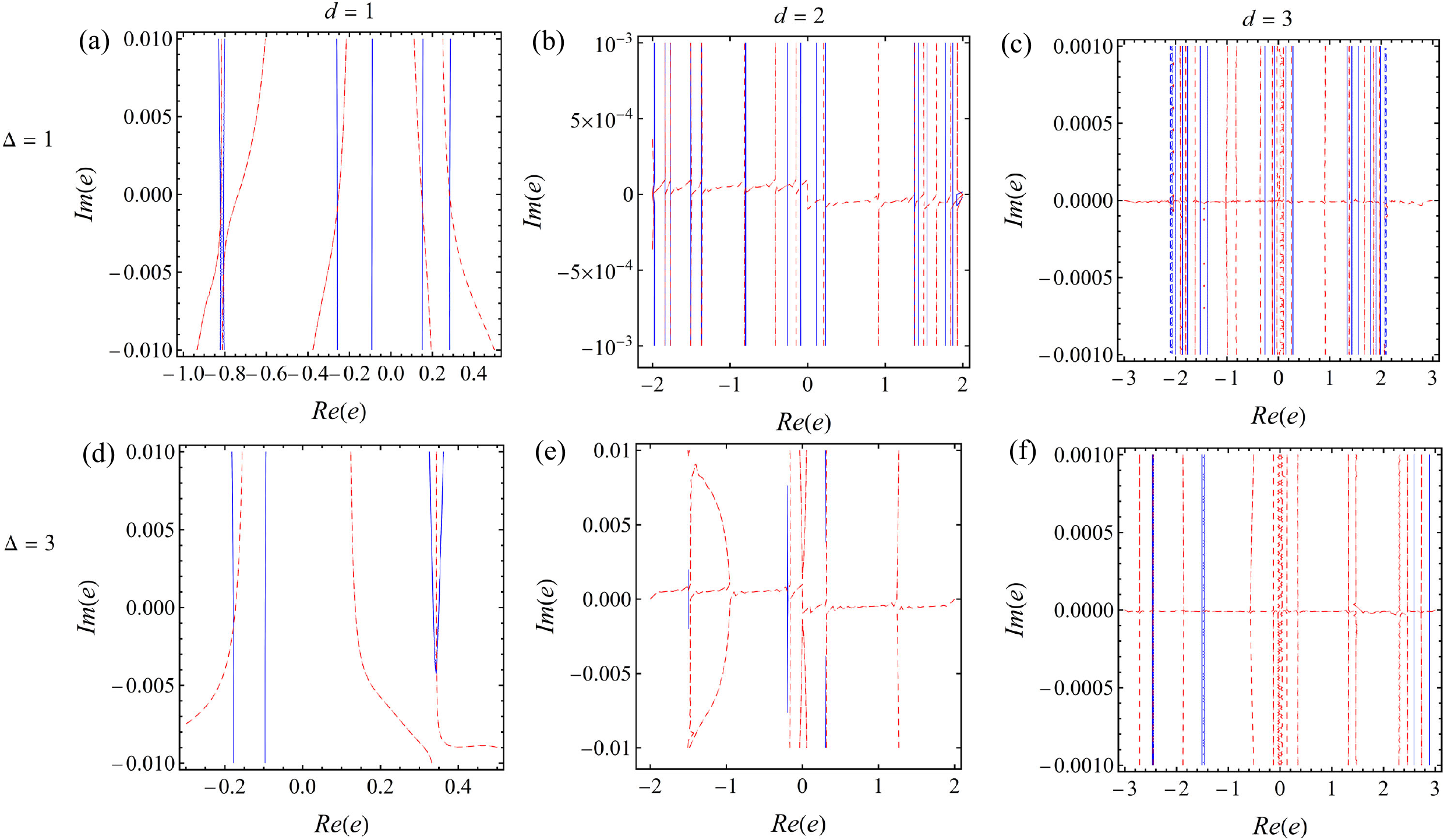}
\caption{(Color online) A detailed  plot for  the zero determinant of the coefficient matrix in Eq. \eqref{seq} close to axis $\text{Im}(e)=0$.  The parameters are selected as the same as those shown  in Fig. \ref{fig:level}}
\label{fig:a1}
\end{figure*}

In Fig. \ref{fig:a1}, a thorough discussion about the  zero determinant of the coefficient matrix in Eq. \eqref{seq} is presented, focusing on the region closed to axis $\text{Im}(e)=0$. It is evident that for $d=1$, the zero points has the imaginary part with the magnitude of order $10^{-3}$. With a increase of $d$, $\text{Im}(e)$ of the zero points becomes very close to be zero. Admittedly, it is possible that the real zero point can appear. However, to verify this point, one has to check the crossing points one by one. Since the purpose of the section II is to provide physical  understanding of the observations in section III, we do not present the calculations in this place.

\begin{figure*}
\center
\includegraphics[width=15cm]{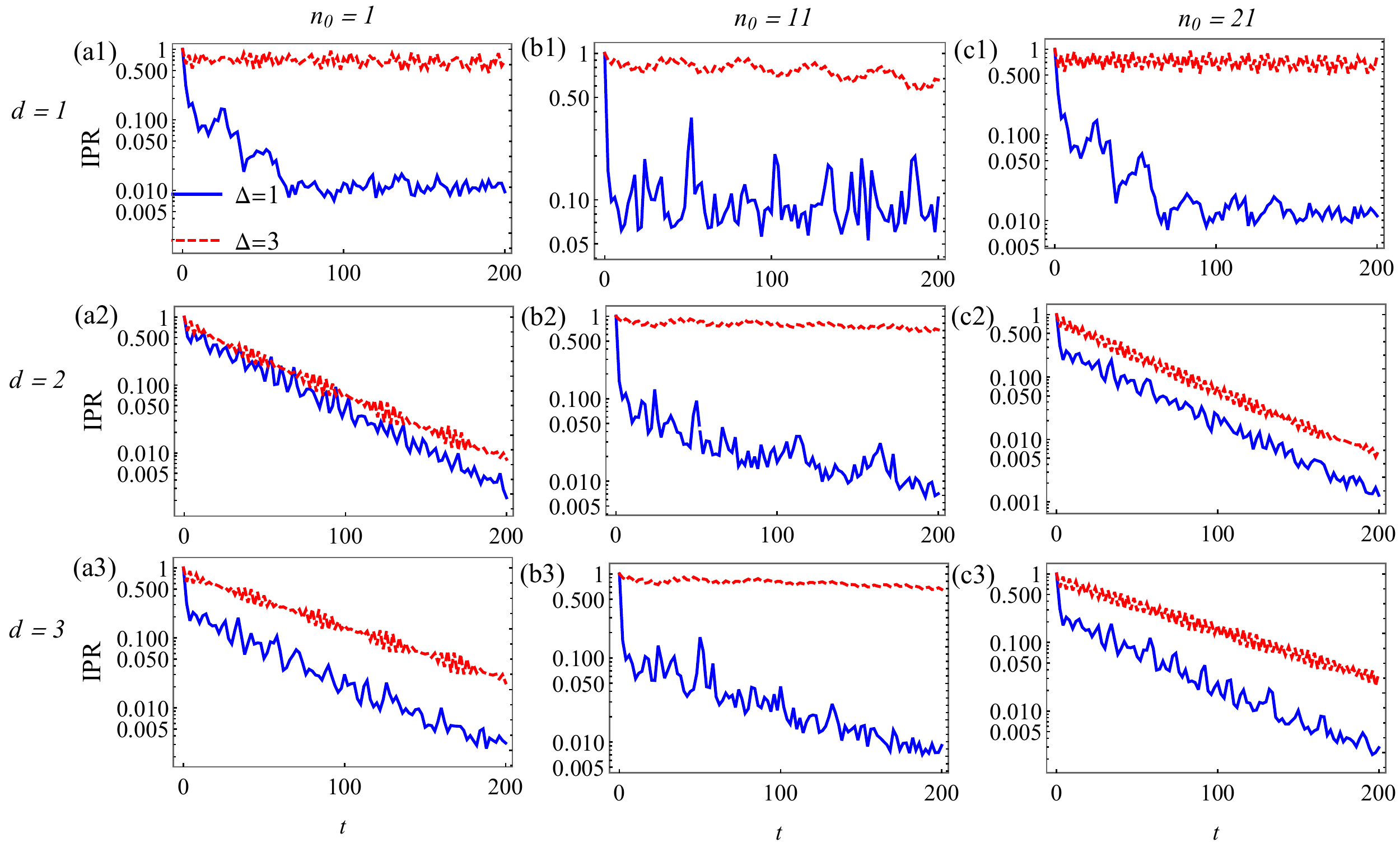}
\caption{(Color online) The logarithm  plots for IPR. The other  parameters are selected as the same as those shown   in Fig. \ref{fig:systemevo}}
\label{fig:ipr}
\end{figure*}

To characterize the distribution of excitation in the atomic sites, the time evolution of  inverse participation ratio (IPR), defined as $\text{IPR}=\sum_{n=1}^N \left|\alpha_n(t)\right|^4$, is plotted in Fig. \ref{fig:ipr}. IPR is a  measure for the localization of state. It has the minimum $1/N$ only if $\left|\alpha_n\right|^2=1/N$ for arbitrary $n$, which means that the distribution of excitation is uniform, and thus the state is extended. However, it has the maximum 1 only if $\left|\alpha_n\right|^2=1$ for a special $n$, which means that excitation  can appear only at site $n$, and thus the state behaves localized strongly. In open case, IPR can also  reflect the extent of retaining the excitation in the system. As shown in Fig. \ref{fig:ipr}, it is found for $d=1$ that IPR  becomes stable around a finite value when $\Delta=1$, which means the finite probability of retaining the excitation in the system. With a increase of $\Delta$, IPR is close to 1. This picture implies that excitation can kept on the initial atomic site  because of the occurrence of bound state. For $d=2$ and $3$, the IPR for initial state  $n_0=1$ and $21$ decays exponentially for whether $\Delta=1$ or $3$. In contrast, the IPR for $n_0=11$ shows a very slow decaying when $\Delta=3$. The underlying reason is the occurrence of bound state, which overlaps with the initial state $n_0=11$ and preserve the information of initial state partially against the decaying into the bath.

\begin{figure*}
\center
\includegraphics[width=15cm]{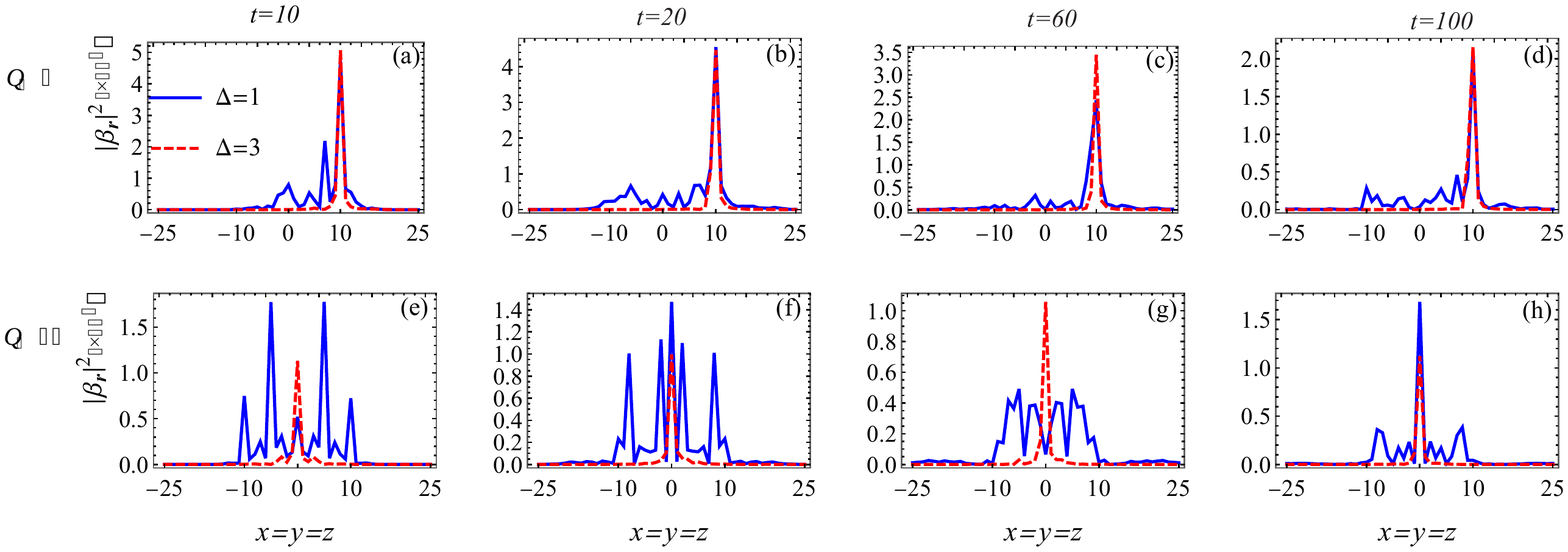}
\caption{(Color online) a  plot for  $\left| \beta_{\mathbf{r}}\right|^2$  along the body-diagonal direction of the cubic lattice bath. The parameters are selected as the same as those shown  in Fig. \ref{fig:3Dconfiger}.}
\label{fig:a3}
\end{figure*}

In Fig. \ref{fig:a3}, a plot for the distribution $\left| \beta_{\mathbf{r}}\right|^2$  along the body-diagonal direction of the cubic lattice is presented  as a complement to plots in Fig. \ref{fig:3Dconfiger}. It is evident that the excitation can hop in the lattice sites coupled to the atomic sites. The spreading of excitation outside these sites is very weak. However, when $\Delta=3$, the excitation can populate significantly on the lattice site coupled to the initial atomic site of excitation. The spreading of excitation in the bath is almost prohibited completely.

\end{document}